\documentclass[11pt]{article}
\usepackage{amsmath}
\usepackage{amssymb}
\usepackage{enumerate}
\usepackage{graphicx}
\usepackage{media9}
\usepackage{color} 
\DeclareMathOperator{\sech}{sech}
\title{Particle Trajectories in Nonlinear Schr\"{o}dinger Models}

\author{John D. Carter\footnote{\texttt{carterj1@seattleu.edu}} \\
{\small Department of Mathematics, Seattle University}
\\ {\small Seattle, WA 98122, USA}
\\\\
Christopher W. Curtis\footnote{\texttt{ccurtis@mail.sdsu.edu}} \\
{\small Department of Mathematics and Statistics, San Diego State University} 
\\ {\small San Diego, CA 92182, USA} 
\\\\
Henrik Kalisch\footnote{\texttt{henrik.kalisch@math.uib.no}}
\\ {\small Department of Mathematics, University of Bergen} 
\\  {\small 5020 Bergen, Norway}
} 

\begin{document}
\maketitle

\begin{center}
{\small
This article is dedicated to Kristian Dysthe on the occasion of his 80th birthday.\\}
\end{center}

\vskip 0.2in

\begin{abstract}
The nonlinear Schr\"{o}dinger equation is well known as a universal equation
in the study of wave motion. In the context of wave motion at the free surface
of an incompressible fluid, the equation accurately predicts the evolution
of modulated wave trains with low to moderate wave steepness.

While there is an abundance of studies investigating the reconstruction of the 
surface profile $\eta$, and the fidelity of such profiles 
provided by the nonlinear Schr\"{o}dinger equation as predictions of real surface water waves, 
very few works have focused on the associated flow field in the fluid.

In the current work, it is shown that the velocity potential $\phi$ can be reconstructed in a similar way as the 
free-surface profile.  This observation opens up a range of potential applications since the nonlinear Schr\"{o}dinger equation features fairly simple closed-form solutions and can be solved numerically with comparatively little effort.  In particular, it is shown that particle trajectories in the fluid can be described with relative ease not only in the context of the nonlinear Schr\"{o}dinger equation, but also in higher-order models such as the Dysthe equation, and in models incorporating certain types of viscous effects.
\end{abstract}

\section{Introduction}
Recent years have seen a flurry of activity aimed at understanding 
the motion of fluid particles in free-surface flows. The problem
has been studied from various point of view, including field 
campaigns, wave-flume experiments, asymptotics, and rigorous mathematical 
analysis.  One of the most well known results about particle trajectories in 
free-surface flows is Stokes's finding that the particle motion
associated to small-amplitude (linear) periodic waves features
a net forward drift which is attributed to the decrease of the
Eulerian particle velocity with increasing depth. 

Recent advances in laboratory technology, in particular those relating to 
particle image velocimitry (PIV) and particle tracking velocimitry (PTV) 
have led to rapid improvements of our understanding of the motion of 
fluid particles in free-surface flow.
The influence of the boundary layers
on the particle drift in a regular wave train was studied in the seminal paper \cite{LH1953}
which was in part inspired by experiments reported in \cite{Bagnold}.
In essence, particle drift will be positive near the bottom and
near the free surface but negative in an intermediate region.
In recent works, experimental measurements are coupled with high-performance data analysis
techniques to paint a rather complete picture of particle motion
in highly nonlinear waves created in a wave flume \cite{GK1,GK2}.
The findings presented in these works are also related to the importance
of the effect of the boundary layer both at the bed and at the free surface.
Indeed, it is observed that the Stokes drift may
take a very different form than what was originally found by Stokes \cite{Stokes}. 
In particular, these recent studies confirm and extend
the results of \cite{LH1953}. 

However, depending on the motion of the wave maker, these results may vary.
For example, the experiments reported in \cite{ChHsCh} seem to confirm
the essence of Stokes's original work. Indeed, 
using both experiments and high-order asymptotics, a strong
case is made in \cite{ChHsCh} that there is a net forward drift throughout the fluid column.
These results are also in line with 
mathematical advances in the understanding
of particle motion in free surface flow (see \cite{EV} for a review). 
In particular, a firm mathematical
proof was given that particle trajectories in Stokes waves are not closed \cite{Co}.

In the case of finite depth, even if viscosity is not taken into
account, mass conservation during the creation phase may lead
to zero drift when averaged over the fluid column \cite{LH1953,RamsdenNath1988}.
On the other hand, in the current analysis, the case of (infinitely) deep water
is investigated and the bottom boundary layer is ignored.
In some studies, in particular in the presence of a background
current in deep water, it has been observed that there is no (Lagrangian) particle
drift either in the average, or even in the pointwise sense \cite{Monismith}.
Some reports of field campaigns also point to the absence of the Stokes
drift in wave trains in the open ocean \cite{Smith}. With these partially conflicting
pieces of evidence, it appears that there is a need for being able to describe
particle paths as accurately as possible using theoretical models.

In the current contribution, we study particle trajectories due to wave motion
described in the narrow-banded spectrum approximation in an idealized infinitely deep fluid.
We focus on irrotational flow which is described in terms of a potential function,
$\phi(x,z,t)$, satisfying Laplace's equation, $\phi_{xx} + \phi_{zz} = 0$, in the fluid domain.
In general terms, if $(\xi(t),\zeta(t))$ denotes the position of a fluid particle in the $(x,z)$-plane 
at a time $t$, then the particle motion is described by the ordinary differential equations 
\begin{subequations}
\begin{equation}
\frac{d\xi}{dt}=\frac{\partial \phi}{\partial
  x}\Big{(}\xi(t),\zeta(t),t\Big{)},
\end{equation}
\begin{equation}
\frac{d\zeta}{dt}=\frac{\partial \phi}{\partial
  z}\Big{(}\xi(t),\zeta(t),t\Big{)}.
\end{equation}
\label{LODEs}
\end{subequations}
The corresponding initial conditions are given by $(\xi(0),\zeta(0))=(\xi_0,\zeta_0)$ 
where $(\xi_0,\zeta_0)$ represents the initial coordinates of the particle in the $(x,z)$-plane.  
In this paper, all variables are dimensional with standard SI units.

As stated above, the focus of this paper is to examine particle trajectories
associated with wave motion which can be approximately described by the nonlinear Schr\"{o}dinger (NLS) 
equation and a few of its generalizations.  This equation arises in the case where a carrier wave of a certain frequency is modulated slowly. In terms of the dimensional slow variables $X = \epsilon x$ and $T = \epsilon t$,
the NLS equation is written as
\begin{equation}
2i \omega_0\Big( B_T+\frac{g}{2\omega_0}B_X\Big{)}+\epsilon\Big{(}\frac{g}{4k_0}B_{XX}+4gk_0^3|B|^2B\Big) = 0,
\label{NLS}
\end{equation}
where $g=9.8 m/s^2$ is the gravitational acceleration,
$B$ describes the dimensional amplitude envelope of the oscillations of the carrier wave,
$k_0>0$ represents the dimensional wave number of the carrier wave, 
$\omega_0 = \sqrt{g k_0}$ represents the dimensional frequency of the carrier wave, 
and $\epsilon=2|a_0|k_0$ represents the (dimensionless) wave steepness.
This equation has been well studied mathematically 
(see, for example, \cite{SulemSulem}) 
and has been shown to favorably predict experimental measurements 
of modulated wave trains when $\epsilon<0.1$ (see for example \cite{LoMei}).

Given a solution, $B(X,T)$, the free surface is reconstructed by
\begin{equation}\label{etaIntro}
\eta(x,t)=\epsilon^3\bar{\eta}+\epsilon
B\mbox{e}^{i\omega_0t-ik_0x}+\epsilon^2B_2\mbox{e}^{2(i\omega_0t-ik_0x)} + \dots + c.c.,
\end{equation}
where $\bar{\eta}$ is a dimensional average term and $B_2$ is the dimensional amplitude of the first harmonic.  These quantities are defined in terms of $B$ and are examined in detail in Section 2.1.  The abbreviation $c.c.$ stands for complex conjugate.  Additionally, in the derivation of the NLS equation, the following ansatz is used for the velocity potential
\begin{equation}\label{phiIntro}
\phi(x,z,t)=\epsilon^2\bar{\phi}+\epsilon A \mbox{e}^{k_0z+i\omega_0t-ik_0x} 
+ \dots + c.c.,
\end{equation}
where $\bar{\phi}=\bar\phi(X,Z,T)$, $A=A(X,Z,T)$, and $Z=\epsilon z$ is the dimensional slow vertical variable.  Substituting this expression into the Laplace equation leads to the following transport equation with imaginary speed
\begin{equation}\label{transport}
A_{Z} - iA_{X}=0,
\end{equation}
at leading order in $\epsilon$.  In addition, at leading order in $\epsilon$ the Bernoulli condition gives
\begin{equation}\label{transportData}
A
=\frac{i\omega_0}{k_0}B,\hspace*{1cm}\text{at $Z=0$.}
\end{equation}
Solving this boundary-value problem for $A$ yields the first-order approximation to the velocity field everywhere in the fluid. It is then possible to compute the particle trajectories associated to any given surface profile numerically.   A number of examples for particle paths that result from NLS solutions are given in the remainder of this section.  Examples of particle paths that result from solutions to higher-order and dissipative generalizations of NLS are examined in the following sections.  Similar considerations regarding particle paths and the potential function have been introduced for a variety of long-wave equations in \cite{AK2,AK4}.  In particular, particle paths and streamlines for waves in the KdV regime were described in detail in \cite{BoKa,He,He3}.  Somewhat different procedures have also recently been used to understand properties of particle motion in the context of the narrow-banded spectrum approximation in the presence of shear flows \cite{CCK} and in the presence of point vortices \cite{CurtisKalisch}.

\subsection{NLS plane-wave solutions}
\begin{figure}
  \begin{center}
  \includegraphics[width=12cm]{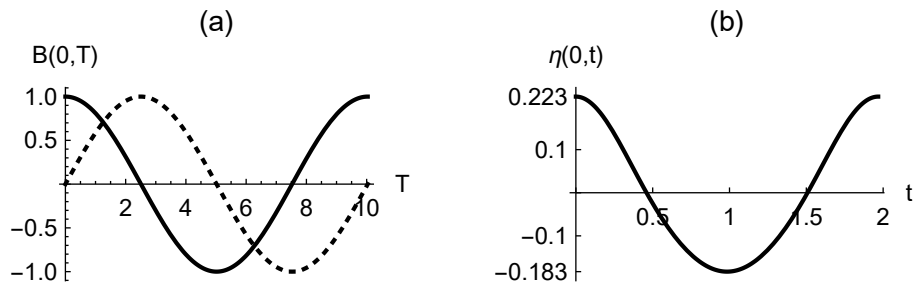}
  \caption{\small Time-series plots of (a) the real (--) and imaginary (- -) parts of the NLS plane-wave solution at $X=0$ and (b) the corresponding surface displacement at $x=0$ for one period of the carrier wave for $\epsilon=0.1$, $k_0=1$, $B_0=1$, and $k=0$.} 
  \label{NLSPWPlots}
  \end{center}
  \end{figure}

The plane-wave solutions of NLS are given by
\begin{equation}
B(X,T)=B_0\mbox{e}^{ikX-i\lambda_{_{NLS}} T},
\label{NLSPWSoln}
\end{equation}
where 
\begin{equation}
\lambda_{_{NLS}}=\Big{(}\frac{k}{2k_0}+\frac{\epsilon
  k^2}{8k_0^2}-2\epsilon B_0^2k_0^2\Big{)}\omega_0,
\end{equation}
and $B_0$ and $k$ are real constants.  Closed-form expressions for the corresponding surface displacement and velocity potential are included in Appendix \ref{etasPWNLS}.  The surface displacement corresponding to this solution has a (temporal) period of 
\begin{equation}
t_{_{NLS}}^*=\frac{2\pi}{\omega_0-\epsilon\lambda_{_{NLS}}}=\frac{16k_0^2\pi}{(8k_0^2-4\epsilon kk_0-\epsilon^2
  k^2+16\epsilon^2B_0^2 k_0^4)\omega_0}.
\label{NLSPeriod}
\end{equation}
For demonstrative purposes, we select $\epsilon=0.1$, $B_0=1$, $k_0=1$ and $k=0$ as parameter values for the plane-wave solution.  For these parameter values, the period of the surface is given by  
$t_{_{NLS}}^*=100\pi\sqrt{5}/357\dot{=}1.968$, 
the crest height is $0.223$, the trough height is $-0.183$, and the wave profile has mean zero.  
The crest and trough heights are not symmetric about $z=0$ even though the mean term is zero because $\eta(x,t)$ is comprised of six (complex) Fourier modes.  See Figure \ref{NLSPWPlots} for time-series plots of one period of this NLS plane-wave solution and one period of the corresponding surface displacement.   

\begin{figure}
\begin{center}
\includegraphics[width=6cm]{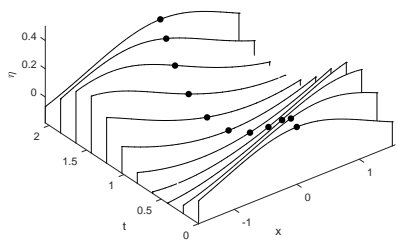}
\caption{\small Evolution of the free surface and a particle located in the surface as described
by a plane-wave solution of the NLS equation with $\epsilon=0.1$ , $k_0 = 1$, $B_0 = 1$, $k = 0$, and
$(\xi_0 , \zeta_0 ) = (0, 0.223)$.
}
\label{NLS3Motion}
\end{center}
\end{figure}
 
\begin{figure}
\begin{center}
\includegraphics[width=12cm]{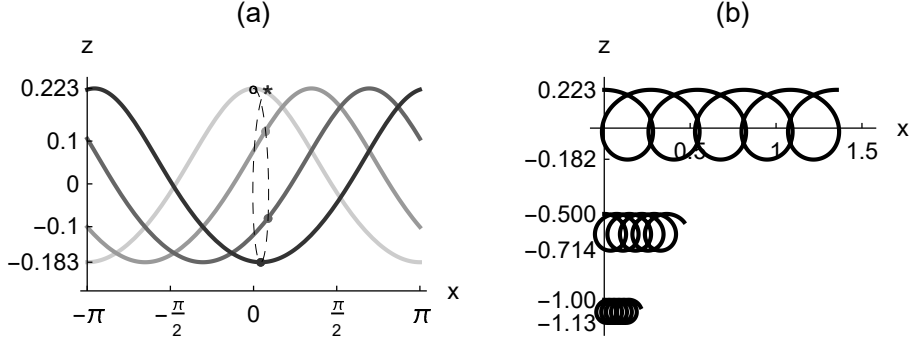}
\caption{\small (a) The fluid surfaces (four solid curves) and positions (dots) of a NLS plane-wave particle that starts on the surface at four equally spaced times in $t\in[0,\frac{1}{2}t_{_{NLS}}(0.223)]$.  The dashed curve represents the path of a particle over an entire period.  The black circle and star represent the initial and final positions of the particle respectively.  (b) The paths of three NLS plane-wave particles with initial positions $(0,0.223)$, $(0,-0.5)$, and $(0,-1)$ on the interval $t\in[0,5t_{_{NLS}}(0.223)]$.}
\label{NLS3LagPlots}
\end{center}
\end{figure}

The paths of NLS plane-wave particles are determined by using \eqref{NLSPWSoln} to determine asymptotic expressions for $\phi$ and $\eta$ that are valid up to $\mathcal{O}(\epsilon^3)$ (the details of this process are described in detail in Section \ref{Construction}) and then numerically integrating the system of ODEs given in equation \eqref{LODEs}.  Figure \ref{NLS3Motion} includes a waterfall plot showing the path of a particle that starts on the surface.  Figure \ref{NLS3LagPlots}(a) shows the fluid surface and the position of a NLS plane-wave particle that starts on the surface at four different $t$ values.  Figure \ref{NLS3LagPlots}(b) shows the paths of three NLS plane-wave particles corresponding to different $\zeta_0$.  The top particle starts on the fluid surface while the other two start (and remain) inside the fluid.

\begin{figure}
\begin{center}
\includegraphics[width=12cm]{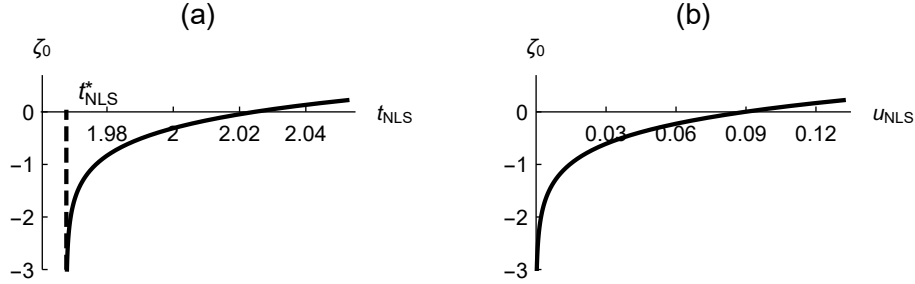}
\caption{\small Plots showing how (a) $t_{_{NLS}}$ and (b) $u_{_{NLS}}$
  change as $\zeta_0$ changes for NLS plane-wave particles with $\epsilon=0.1$, $k_0=1$, $B_0=1$, and $k=0$.}
\label{NLS3LagData}
\end{center}
\end{figure}

Let $t_{_{NLS}}(\zeta_0)$ represent the period of the vertical motion of a particle 
with initial $z$ coordinate $\zeta_0$.  
Values for $t_{_{NLS}}(\zeta_0)$ were computed by numerically solving equation \eqref{LODEs} 
and determining the period of the resulting solution for a range of $\zeta_0$ values.  
A plot showing how $t_{_{NLS}}(\zeta_0)$ and $\zeta_0$ are related is included in Figure \ref{NLS3LagData}(a).  
Note that
\begin{subequations}
\begin{equation}
t_{_{NLS}}(\zeta_0)>t_{_{NLS}}^*,\hspace*{1cm}\text{when $\zeta_0>-\infty$},
\end{equation}
\begin{equation}
\lim_{\zeta_0\rightarrow-\infty}t_{_{NLS}}(\zeta_0)=t_{_{NLS}}^*.
\end{equation}
\end{subequations}
The horizontal motion of the particles is not periodic because $\xi(t_{_{NLS}}(\zeta_0))>\xi_0$ regardless of the initial position of the particle.  (Note that this does not mean the particle always moves to the right.)  The difference between the final and initial horizontal positions during one period of vertical motion,
$\xi(t_{_{NLS}}(\zeta_0))-\xi_0$, is known as the horizontal Lagrangian drift.  Figure \ref{NLS3LagData}(b) contains a plot relating the average horizontal Lagrangian velocity,
\begin{equation}
u_{_{NLS}}(\zeta_0)=\frac{\xi(t_{_{NLS}}(\zeta_0))-\xi_0}{t_{_{NLS}}(\zeta_0)},
\label{LagVel}
\end{equation}
and $\zeta_0$.  As expected from equation \eqref{BottomBC}, 
both $u_{_{NLS}}(\zeta_0)$ and the horizontal Lagrangian drift limit to zero as $\zeta_0\rightarrow-\infty$.

Additional simulations (not shown) establish that increasing $B_0$ increases $t_{_{NLS}}$, $u_{_{NLS}}$ and the horizontal Lagrangian drift in roughly exponential manners.  Increasing the wave-number of the solution, $k$, increases $t_{_{NLS}}$, and decreases $u_{_{NLS}}$ and the drift in approximately linear manners.

\begin{figure}
\begin{center}
\includegraphics[width=6cm]{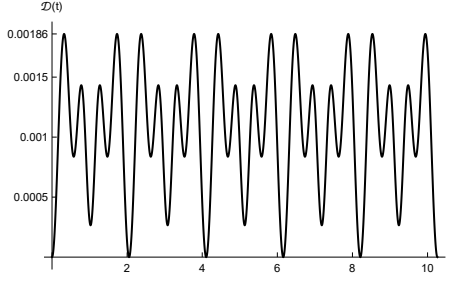}
\caption{\small A plot of $\mathcal{D}(t)$ for a NLS plane-wave solution with $\epsilon=0.1$, $B_0=1$, $k_0=1$ and $k=0$ on $t\in[0,5t_{_{NLS}}(0.223)]$.}
\label{NLS3Diff}
\end{center}
\end{figure}

Equation \eqref{BC1}, the kinematic boundary condition, implies that a particle that starts on the surface stays on the surface.  This equation is only approximately satisfied by NLS since NLS is an asymptotic approximation to the Euler equations.  Therefore NLS particles that start on the surface of the fluid do not necessarily remain on the surface.  A plot of the difference between the particle's vertical position and the fluid surface,
\begin{equation}
\mathcal{D}(t)=\big{|}\zeta(t)-\eta\big{(}\xi(t),t\big{)}\big{|},
\label{Diff}
\end{equation}
is included in Figure \ref{NLS3Diff}.  As an additional check on our
work, we computed
\begin{equation}
\mathcal{E}(\epsilon)=\int_0^{t_{_{NLS}}}\mathcal{D}(t)~dt,
\label{error}
\end{equation}
for a variety of values of $\epsilon\in(0,\frac{1}{4}]$.  Table
\ref{NLSErrors} contains a summary of these results that establish $\mathcal{E}(\epsilon)\sim\mathcal{O}(\epsilon^4)$ for plane-wave solutions of NLS (as expected because NLS is an $\mathcal{O}(\epsilon^3)$ approximation).

\begin{table}
\begin{center}
\begin{tabular}{|c c c c|}
\hline
$j$ & $\epsilon$ & $\mathcal{E}_j$ &
$\frac{\mathcal{E}_j}{\mathcal{E}_{j-1}}$ \\
\hline
1 & $1/4$ & $6.05*10^{-2}$ & ---\\
2 & $1/8$ & $2.89*10^{-3}$ & 20.9\\
3 & $1/16$ & $1.51*10^{-4}$ & 19.1\\
4 & $1/32$ & $8.78*10^{-6}$ & 17.2\\
6 & $1/64$ & $5.32*10^{-7}$ & 16.5\\
7 & $1/128$ & $3.27*10^{-8}$ & 16.2\\
8 & $1/256$ & $2.03*10^{-9}$ & 16.1\\
\hline
\end{tabular}
\caption{\small Table of $\mathcal{E}(\epsilon)$ for eight values of
  $\epsilon = 1/2^{j+1}$ for the plane-wave solutions of NLS with $k_0=1$, $B_0=1$, and $k=0$.}
\label{NLSErrors}
\end{center}
\end{table}

\subsection{Cnoidal-wave solutions of NLS}

\begin{figure}
\begin{center}
\includegraphics[width=12cm]{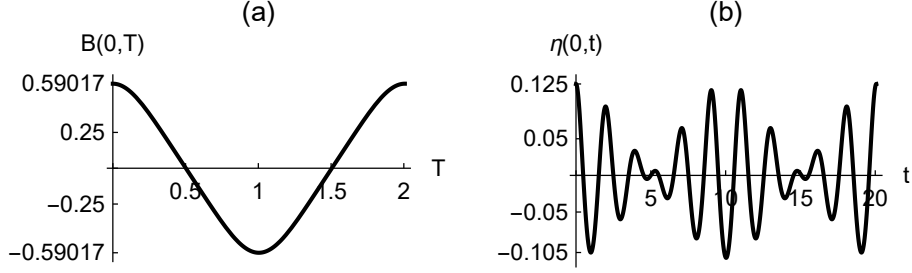}
\caption{\small Time-series plots of one period of the NLS cnoidal-wave solution and (b) one period of the corresponding surface displacement for $\kappa=1/\sqrt{2}$, $\epsilon=0.1$, $k_0=1$, and $B_0=K(1/\sqrt{2})/\pi$. Note that the imaginary part of $B$ is zero in this case.} 
\label{NLSCN1Plots}
\end{center}
\end{figure}

\begin{figure}
\begin{center}
\includegraphics[width=6cm]{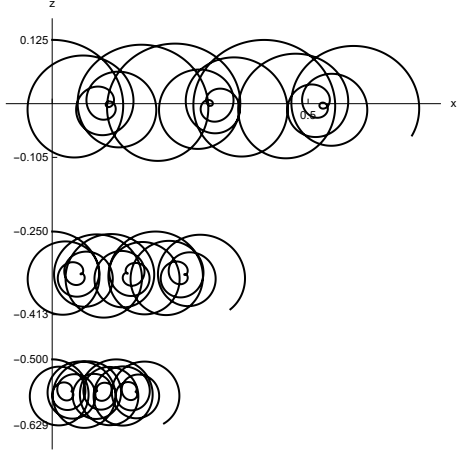}
\caption{\small A plot showing the paths of three particles at different levels on the interval $t\in [0, 30]$ during the propagation of a NLS cnoidal-wave solution with $\kappa=1/\sqrt{2}$, $\epsilon=0.1$, $k_0=1$, and
$B_0=K(1/\sqrt{2})/\pi$.  The initial particle positions are $(0,0.1248)$, $(0,-0.25)$, and $(0,-0.5)$.}
\label{CNLagPP1}
\end{center}
\end{figure}

\begin{figure}
\begin{center}
\includegraphics[width=12cm]{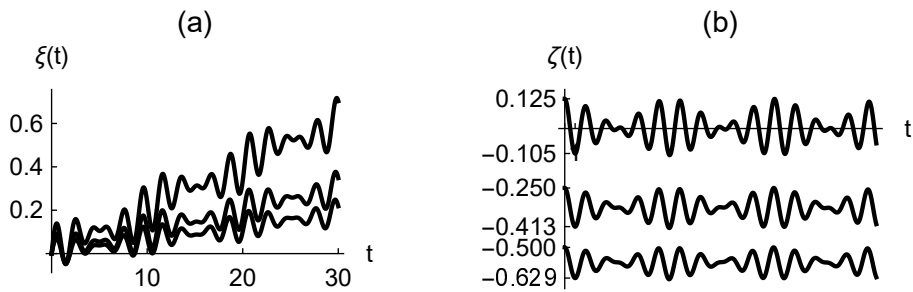}
\caption{\small Plots of (a) $\xi(t)$ and (b) $\zeta(t)$ for the three NLS
  cnoidal-wave particles in Figure \ref{CNLagPP1}.}
\label{CNxPlot1}
\end{center}
\end{figure}

The cnoidal-wave solutions of NLS are given by
\begin{equation}
B(X,T)=B_0~\mbox{cn}\Big{(}\frac{2\sqrt{2}~B_0k_0^2}{\kappa}\big{(}X-\frac{\omega_0}{2k_0}T\big{)},\kappa\Big{)}\mbox{e}^{i\epsilon
  B_0^2k_0^2\omega_0(2\kappa^2-1)T/\kappa^2},
\label{cnsoln}
\end{equation}
where $B_0$ and $\kappa\in[0,1)$ are real parameters.  Here $\mbox{cn}(\cdot,\kappa)$ is a Jacobi elliptic function with elliptic modulus $\kappa$ and period $4K(\kappa)$ where $K$ is the complete elliptic integral of the first kind~\cite{bf}.  Formulas for the corresponding surface displacement and velocity potential are included in Appendix \ref{etasCNNLS}.   There are two situations in which $B$ is periodic in $T$.  First, if $\kappa=1/\sqrt{2}$, then $B$ is real for all $X$ and $T$ and has a $T$-period of $\frac{2K(1/\sqrt{2})}{B_0k_0\omega_0}$.  In this case, if $B_0=\frac{K(1/\sqrt{2})}{\pi k_0}$, then the solution corresponds to a periodic surface displacement with $t$-period
\begin{equation}
t_{_{CN}}^*=\frac{2\pi}{\epsilon\omega_0}.
\end{equation}
Plots of the NLS cnoidal-wave solution and the corresponding surface displacement for $\kappa=1/\sqrt{2}$, $\epsilon=0.1$, $k_0=1$, and $B_0=K(1/\sqrt{2})/\pi$ are included in Figure \ref{NLSCN1Plots}.  Plots of the paths of three particles are included in Figures \ref{CNLagPP1} and \ref{CNxPlot1}.  Unlike plane-wave solutions, cnoidal-wave solutions do not have constant magnitude.  This means that the vertical motion of the particles (except in rare cases) is quasiperiodic rather than periodic.  For example, the particle that starts at a peak on the surface has an initial elevation of $\zeta_0=0.1248$.  The vertical motion of this particle has a quasiperiod of $t=20.2047$.  After one quasiperiod, the particle reaches an elevation of $\zeta(20.2047)=0.1246$, slightly lower than its initial height.  This is caused by the fact that the particle experienced a horizontal drift of 0.4140 during this interval.  Due to this quasiperiodicity, the Lagrangian drift and average Lagrangian velocity are not well defined.

\begin{figure}
\begin{center}
\includegraphics[width=12cm]{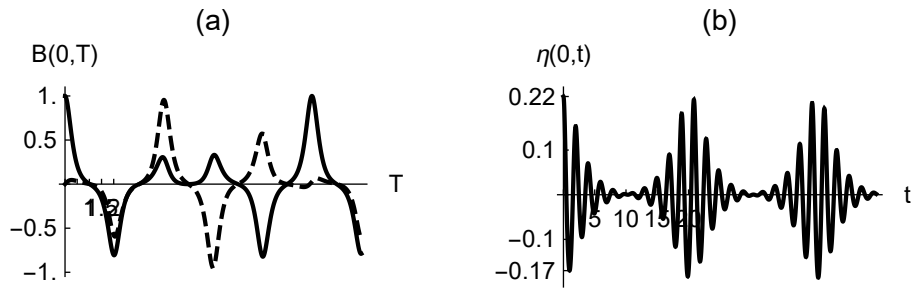}
\caption{\small Time-series plots of (a) the real (--) and imaginary (- -) parts of the cnoidal-wave solution to NLS and (b) the corresponding surface displacement corresponding to $\kappa=0.999$, $\epsilon=0.1$, $k_0=1$, and $B_0=1$.} 
\label{NLSCN2Plots}
\end{center}
\end{figure}

\begin{figure}
\begin{center}
\includegraphics[width=6cm]{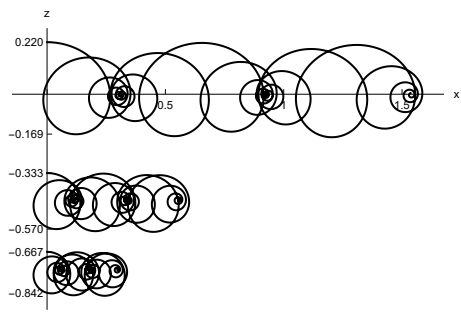}
\caption{\small A plot showing how three NLS cnoidal-wave solutions with
  $\epsilon=0.1$, $\kappa=0.999$, $k_0=1$, and $B_0=1$ evolve
  over $t\in[0,50]$.  The initial positions are $(0,0.220)$,
  $(0,-0.333)$, and $(0,-0.666)$.}
\label{CNLagPP2}
\end{center}
\end{figure}

\begin{figure}
\begin{center}
\includegraphics[width=12cm]{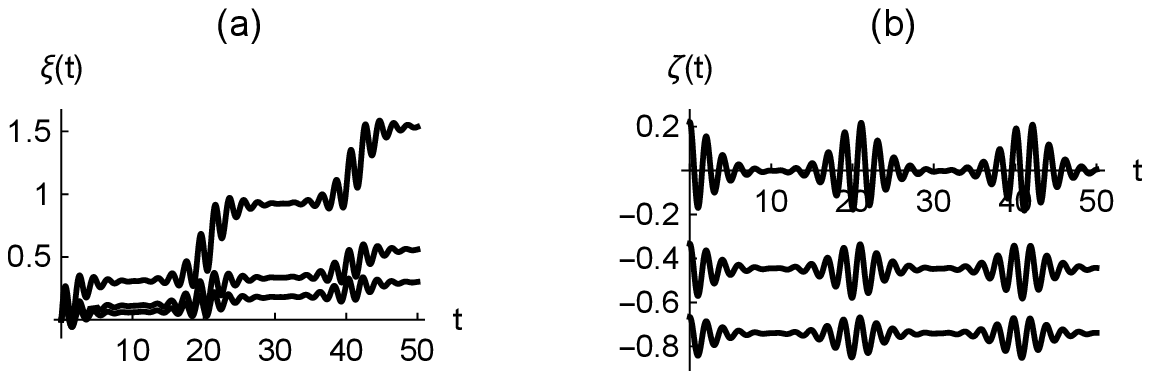}
\caption{\small Plots of (a) $\xi(t)$ and (b) $\zeta(t)$ for the three NLS
  cnoidal-wave particles in Figure \ref{CNLagPP2}.}
\label{CNxPlot2}
\end{center}
\end{figure}

The second situation in which a $T$-periodic solution is obtained is if $B_0$ and $\kappa\ne\frac{1}{\sqrt{2}}$ satisfy
\begin{equation}
B_0=\frac{\sqrt{2}~\kappa\pi}{\epsilon(2\kappa^2-1)k_0K(\kappa)}.
\end{equation}
In order for this solution to correspond to a $t$-periodic surface displacement, the period of this solution must align with the period of the carrier wave.  Enforcing this restriction leads to large-amplitude surface displacements that are well outside of the regime where NLS is expected to be valid.  Therefore, for demonstrative purposes, we select $\epsilon=0.1$, $k_0=1$, $\kappa=0.999$ and $B_0=1$ (parameters that lead to a non-periodic surface displacement in the NLS regime).  Plots of this solution and the corresponding surface displacement are included in Figure \ref{NLSCN2Plots}.  Figures \ref{CNLagPP2} and \ref{CNxPlot2} show how three of these NLS cnoidal-wave particles move in $t$.  Just as in the previous case, the vertical motion of the particles is quasiperiodic even though the solution is periodic.

For the cnoidal-wave solutions of NLS, increasing $B_0$ increases the horizontal drift for particles that start on the surface.  

\subsection{Solitary-wave solutions of NLS}
\begin{figure}
\begin{center}
\includegraphics[width=12cm]{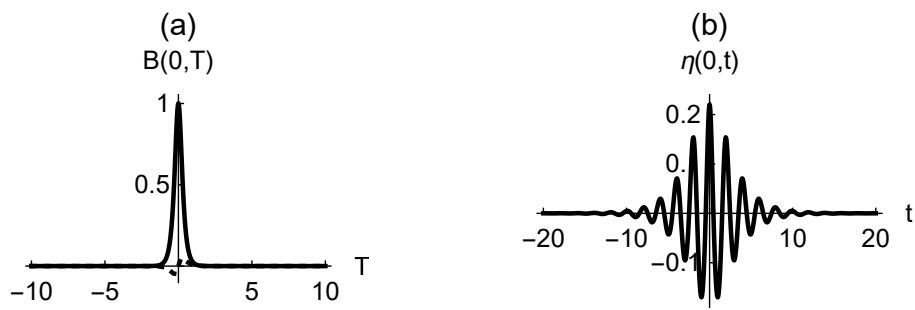}
\caption{\small Time-series plots of (a) the real (--) and imaginary (- -) parts of the solitary-wave solution to NLS and (b) the corresponding surface displacement for $\epsilon=0.1$, $k_0=1$, and $B_0=1$.} 
\label{NLSSechPlots} 
\end{center}
\end{figure}

\begin{figure}
\begin{center}
\includegraphics[width=6cm]{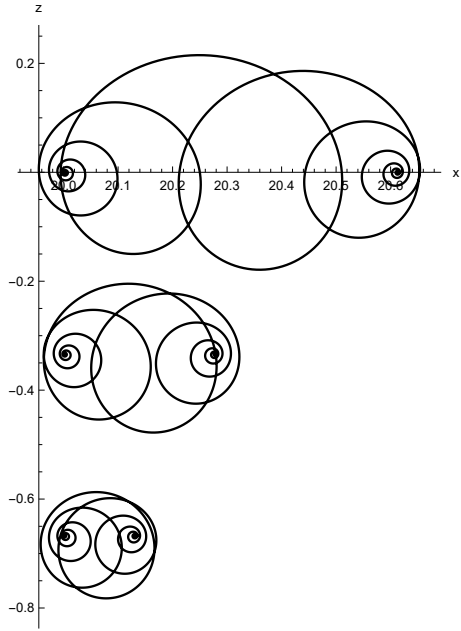}
\caption{\small A plot showing the paths of three particles at different levels on the interval $t\in [0, 50]$ during the propagation of a NLS solitary-wave solution with $\epsilon=0.1$, $k_0=1$, and $B_0=1$.  The initial particle positions are $(20,6.33*10^{-4})$ (a point on the surface), $(20,-0.333)$, and $(20,-0.666)$.}
\label{SechLagPP}
\end{center}
\end{figure}

\begin{figure}
\begin{center}
\includegraphics[width=12cm]{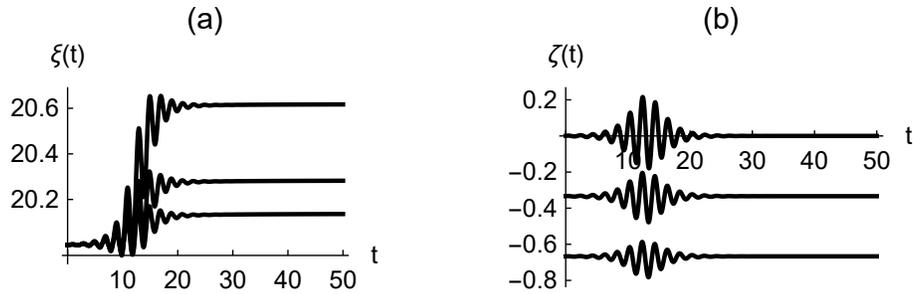}
\caption{\small Plots of (a) $\xi(t)$ and (b) $\zeta(t)$ for the three NLS solitary-wave particles in Figure \ref{SechLagPP}.}
\label{SechxPlot}
\end{center}
\end{figure}

The solitary-wave solutions of NLS,
\begin{equation}
B(X,T)=B_0~\mbox{sech}\Big{(}2\sqrt{2}~B_0k_0^2\big{(}X-\frac{\omega_0}{2k_0}T\big{)}\Big{)}\mbox{e}^{i\epsilon
  B_0^2k_0^2\omega_0T},
\label{sechsoln}
\end{equation}
are obtained from the cnoidal-wave solutions via the limit $\kappa\rightarrow1^-$.  Formulas for the corresponding surface displacement and velocity potential are included in Appendix \ref{etasSechNLS}.  This solution is not
periodic in $X$ or $T$ and therefore does not lead to a periodic surface displacement.   We select the parameter values $\epsilon=0.1$, $k_0=1$, and $B_0=1$ for demonstrative purposes.  
Figure \ref{NLSSechPlots} contains plots of this solution and the corresponding surface displacement.  
Figures \ref{SechLagPP} and \ref{SechxPlot} show the paths of three NLS solitary-wave particles on the interval $t\in[0,50]$.  During $t\in[0,50]$, the horizontal drift for the particle that starts on the surface is approximately $0.6170$.

In the next section, we present the full derivation of the velocity potential
in the more general case of the higher-order NLS (Dysthe) equation
and the viscous models. Then in Section 3, we describe particle paths 
for a number of examples.

\section{Construction of the velocity potential}
\label{Construction}
We now give the full details of the reconstruction of the velocity potential
in terms of the unknown $B(X,T)$ of the nonlinear Schr\"{o}dinger equations.
In fact, we will take a slightly more general view by including viscous effects,
and higher-order terms.  In order to explain how to find the potential $\phi$,
it is convenient to first review the derivation of the evolution
equation.

\subsection{Derivation of the Viscous Dysthe System}
Wu {\emph{et al.}}~\cite{WuLiuYue} proposed the following system for a two-dimensional, infinitely-deep, weakly-dissipative fluid
\begin{subequations}\label{DDZ}
\begin{equation}
\phi_{xx}+\phi_{zz}=0, \hspace*{1cm}\mbox{for } -\infty<z<\eta,
\label{Laplace}
\end{equation}
\begin{equation}
\phi_t+\frac{1}{2}\big{(}\phi_{x}^2+\phi_{z}^2\big{)}+g\eta=-4\bar{\alpha}\phi_{zz},\hspace*{1cm}\mbox{at
} z=\eta,
\label{BC2}
\end{equation}
\begin{equation}
\eta_t+\eta_x\phi_x=\phi_z,\hspace*{1cm}\mbox{at } z=\eta,
\label{BC1}
\end{equation}
\begin{equation}
\phi_x,~\phi_z\rightarrow0, \hspace*{1cm}\mbox{as } z\rightarrow-\infty.
\label{BottomBC}
\end{equation}
\end{subequations}
Here $\phi=\phi(x,z,t)$ represents the velocity potential of the fluid, $\eta=\eta(x,t) $ represents the free-surface displacement, $g$ represents the acceleration due to gravity, and $\bar{\alpha}>0$ represents dissipation from all sources.  The classical Euler equations (see, for example, Debnath~\cite{Debnath}) are recovered from this system by setting $\bar{\alpha}=0$. Following the work of Zakharov~\cite{zak68} and Dysthe~\cite{Dysthe}, 
we make the following modulated wave-train ansatz
\begin{subequations}
\begin{equation}
\eta(x,t)=\epsilon^3\bar{\eta}+\epsilon
B\mbox{e}^{i\omega_0t-ik_0x}+\epsilon^2B_2\mbox{e}^{2(i\omega_0t-ik_0x)}
+\dots+c.c.,
\label{etaansatz}
\end{equation}
\begin{equation}
\phi(x,z,t)=\epsilon^2\bar{\phi}+\epsilon
A_1\mbox{e}^{k_0z+i\omega_0t-ik_0x}+\epsilon^2
A_2\mbox{e}^{2(k_0z+i\omega_0t-ik_0x)}
+\dots+c.c.,
\label{phiansatz}
\end{equation}
\end{subequations}
where $\omega_0$ represents the frequency of the carrier wave, $k_0>0$ represents the wave number of the carrier wave, $\epsilon=2|a_0|k_0$ represents the (dimensionless) wave steepness, $a_0$ represents a typical amplitude, and $c.c.$ represents complex conjugate. Further assume
\begin{subequations}
\begin{equation}
\bar{\eta}=\bar{\eta}(X,T)=\bar{\eta}_0(X,T)+\epsilon \bar{\eta}_1(X,T)+\epsilon^2
\bar{\eta}_2(X,T)+\cdots,
\end{equation}
\begin{equation}
B=B(X,T),
\end{equation}
\begin{equation}
B_j=B_j(X,T)=B_{j0}(X,T)+\epsilon B_{j1}(X,T)+\epsilon^2
B_{j2}(X,T)+\cdots,\hspace*{0.5cm}\text{for $j=2,3,\dots$},
\end{equation}
\begin{equation}
\bar{\phi}=\bar{\phi}(X,Z,T)=\bar{\phi}_0(X,Z,T)+\epsilon \bar{\phi}_1(X,Z,T)+\epsilon^2
\bar{\phi}_2(X,Z,T)+\cdots.
\end{equation}
\begin{equation}
A_j=A_j(X,Z,T)=A_{j0}(X,Z,T)+\epsilon A_{j1}(X,Z,T)+\epsilon^2
A_{j2}(X,Z,T)+\cdots,\hspace*{0.5cm}\text{for $j=1,2,\dots$}.
\end{equation}
\end{subequations}
The slow space and time variables are defined by $X=\epsilon x$, $Z=\epsilon z$, and $T=\epsilon t$.  
Assuming that $\bar{\alpha}=\epsilon^2\alpha$ and carrying out the perturbation analysis 
through $\mathcal{O}(\epsilon^4)$ leads to (see \cite{FD,FD2} for details)
\begin{subequations}\label{vDysthe1}
\begin{equation}
\begin{split}
2i&\omega_0\Big{(}B_T+\frac{g}{2\omega_0}B_X\Big{)}+\epsilon\Big{(}\frac{g}{4k_0}B_{XX}+4gk_0^3|B|^2+4ik_0^2\omega_0\alpha
B\Big{)}
+\epsilon^2 \Big{(}-\frac{ig}{8k_0^2}B_{XXX}
\\ &
+2igk_0^2B^2B_X^*+12igk_0^2|B|^2B_X+2k_0\omega_0B\bar{\phi}_{0X}-8k_0\omega_0\alpha
B_X\Big{)}=0,\hspace*{0.5cm}\text{at $Z=0$},
\end{split}
\end{equation}
\begin{equation}
\bar{\phi}_{0Z}=2\omega_0\big{(}|B|^2\big{)}_X,\hspace*{1cm}\text{at }Z=0,
\end{equation}
\begin{equation}
\bar{\phi}_{0XX}+\bar{\phi}_{0ZZ}=0,\hspace*{1cm}\text{for }-\infty<Z<0,
\end{equation}
\begin{equation}
\bar{\phi}_{0X},~\bar{\phi}_{0Z}\rightarrow0,\hspace*{1cm}\text{as }Z\rightarrow-\infty,
\end{equation}
\end{subequations}
where $\omega_0^2=gk_0$ and $B^*$ represents the complex conjugate of $B$.  If the Hilbert transform, $\mathcal{H}$, is defined by
\begin{equation*}
\mathcal{H}\big{(}f(x)\big{)}=\mathcal{F}^{-1}\big{(}-i\mbox{sgn}(k)\hat{f}(k)\big{)},
\end{equation*}
where the Fourier transform and its inverse are defined by
\begin{equation*}
\hat{f}(k)=\mathcal{F}\big{(}f(x)\big{)}=\int_\mathbb{R} f(x)\mbox{e}^{-ikx}dx,\hspace*{1cm}f(x)=\mathcal{F}^{-1}\big{(}\hat{f}(k)\big{)}=\frac{1}{2\pi}\int_\mathbb{R} \hat{f}(k)\mbox{e}^{ikx}dk, 
\end{equation*}
then the system \eqref{vDysthe1} can be rewritten as
\begin{equation}
\begin{split}
&2i\omega_0\Big{(}B_T+\frac{g}{2\omega_0}B_X\Big{)}+\epsilon\Big{(}\frac{g}{4k_0}B_{XX}+4gk_0^3|B|^2B+4ik_0^2\omega_0\alpha
B\Big{)}\\ & +\epsilon^2 \Big{(} \!- \!\frac{ig}{8k_0^2}B_{XXX} \!+ \! 2igk_0^2B^2B_X^*+12igk_0^2|B|^2B_X \!- \! 
4gk_0^2B\big{(}\mathcal{H}(|B|^2)\big{)}_X \!- \!8k_0\omega_0\alpha
B_X\Big{)}=0,
\end{split}
\label{vDysthe}
\end{equation}
with
\begin {equation*}
\bar{\phi}_0(X,Z,T)=\frac{\omega_0}{\pi}\int_\mathbb{R}i\mbox{sgn}(k)\mathcal{F}\Big{(}|B|^2\Big{)}\mbox{e}^{ikx+|k|z}dk.
\end {equation*}
We refer to this system as the viscous Dysthe (vDysthe) system.  
The NLS equation is obtained from this system by setting $\alpha=0$ 
and disregarding terms of order $\epsilon^2$ and higher.  
The Dysthe system, also known as the modified NLS system, is obtained from Equation \eqref{vDysthe} 
by setting $\alpha=0$.  The dissipative NLS equation (dNLS) is obtained by setting $\epsilon^2=0$.

Given a solution to Equation \eqref{vDysthe}, obtaining the corresponding surface displacement is straightforward but tedious.  Determining the velocity potential is more complicated (and more tedious) because partial differential equations (PDEs) must be solved to determine each $A_{jk}$.  To this order, 
the surface displacement and velocity potential are given by (see Section \ref{AsymptoticDetail} for more detail)
\begin{subequations}\label{surfvelpot}
\begin{equation}
\begin{split}
\eta(x,t) 
= & \, \epsilon BE+\epsilon^2k_0B^2E^2 + \epsilon^3 \Big(\frac{3}{2}k_0^2B^3E^3+iBB_XE^2+\frac{2\omega_0}{g}\big{(}\mathcal{H}(|B|^2)\big{)}_T\Big)\\
 & + \epsilon^4 \Big(\frac{4ik_0^2\omega_0\alpha}{g}B^2E^2+\frac{17k_0^3}{3}|B|^2B^2E^2+\frac{8k_0^3}{3}B^4E^4+3ik_0B^2B_XE^3-\frac{1}{4k_0}|B_X|^2\\
&-\frac{1}{4k_0}BB^*_{XX}-\frac{1}{g}\bar{\phi}_{1T}\Big)+\mathcal{O}(\epsilon^5)+c.c.,
\end{split}
\label{eta}
\end{equation}
\begin{equation}
\begin{split}
\phi(x,z,t)=&\bigg\{\frac{i\epsilon\omega_0}{k_0}\check{B}+\frac{\epsilon^2\omega_0}{2k_0^2}\check{B}_X+\epsilon^3\Big(-\frac{ik_0\omega_0}{2}|\check{B}|^2\check{B}-\frac{3i\omega_0}{8k_0^3}\check{B}_{XX}\Big)\\&
+\epsilon^4\Big(-\frac{\omega_0}{4}\check{B}^2\check{B}^*_X+\frac{\omega_0}{2}|\check{B}|^2\check{B}_X-\frac{5\omega_0}{16k_0^4}\check{B}_{XXX}-2ik_0\check{B}\big{(}\mathcal{H}(|\check{B}|^2)\big{)}_T\Big)\bigg\}E\mbox{e}^{k_0z}\\& +\epsilon^4\Big{(}-2k_0^2\alpha\check{B}^2+4ik_0^2\omega_0|\check{B}|^2\check{B}^2\Big{)}E^2\mbox{e}^{2k_0z}+\mathcal{O}(\epsilon^5)+c.c.,
\label{phi}
\end{split}
\end{equation}
\end{subequations}
\noindent
where $\check{B}=B(X+iZ,T)$, $\check{B^*}=B^*(X+iZ,T)$, $E=\mbox{e}^{i\omega_0t-ik_0x}$,
and $c.c.$ represents the complex conjugate.
Unfortunately, a simple, general formula for $\bar{\phi}_1$ does not exist.  
However, we include formulas for $\bar{\phi}_1$ for the particular solutions we examine below.

\subsection{Derivation of the velocity potential}
\label{AsymptoticDetail}
In order to determine the $Z$ dependence of each of the terms in the expansion for the velocity potential, PDEs must be solved at each order.  At $\mathcal{O}(\epsilon)$ the PDE and surface boundary condition are
\begin{subequations}
\begin{equation}
A_{10Z} - iA_{10X}=0,
\end{equation}
\begin{equation}
A_{10}=\frac{i\omega_0}{k_0}B,\hspace*{1cm}\text{at $Z=0$.}
\end{equation}
\end{subequations}
At $\mathcal{O}(\epsilon^2)$ the PDEs and surface boundary conditions are
\begin{subequations}
\begin{equation}
A_{20Z}-iA_{20X}=0,
\end{equation}
\begin{equation}
A_{11Z}-iA_{11X}=-\frac{1}{2k_0}\Delta A_{10},
\end{equation}
\begin{equation}
A_{20}=0,\hspace*{1cm}\text{at $Z=0$,}
\end{equation}
\begin{equation}
A_{11}=\frac{\omega_0}{2k_0^2}B_X,\hspace*{1cm}\text{at $Z=0$.}
\end{equation}
\end{subequations}
At $\mathcal{O}(\epsilon^3)$ the PDEs and surface boundary conditions are
\begin{subequations}
\begin{equation}
A_{30Z}-iA_{30X}=0,
\end{equation}
\begin{equation}
A_{21Z}-iA_{21X}=-\frac{1}{4k_0}\Delta A_{20},
\end{equation}
\begin{equation}
A_{12Z}-iA_{12X}=-\frac{1}{2k_0}\Delta A_{11},
\end{equation}
\begin{equation}
A_{30}=0,\hspace*{1cm}\text{at $Z=0$,}
\end{equation}
\begin{equation}
A_{21}=0,\hspace*{1cm}\text{at $Z=0$,}
\end{equation}
\begin{equation}
A_{12}=-2k_0\alpha B+\frac{3}{2}ik_0\omega_0|B|^2B-\frac{i\omega_0}{4k_0^3}B_{XX},\hspace*{1cm}\text{at $Z=0$.}
\end{equation}
\end{subequations}
At $\mathcal{O}(\epsilon^4)$
\begin{subequations}
\begin{equation}
A_{40Z}-iA_{40X}=0
\end{equation}
\begin{equation}
A_{31Z}-iA_{31X}=-\frac{1}{6k_0}\Delta A_{30}
\end{equation}
\begin{equation}
A_{22Z}-iA_{22X}=-\frac{1}{4k_0}\Delta A_{21}
\end{equation}
\begin{equation}
A_{13Z}-iA_{13X}=-\frac{1}{2k_0}\Delta A_{12}
\end{equation}
\begin{equation}
A_{40}=0,\hspace*{1cm}\text{at $Z=0$,}
\end{equation}
\begin{equation}
A_{31}=0,\hspace*{1cm}\text{at $Z=0$,}
\end{equation}
\begin{equation}
A_{22}=-6k_0^2\alpha B^2+8ik_0^2\omega_0|B|^2B^2+\frac{i\omega_0}{4k_0^2}BB_{XX},\hspace*{1cm}\text{at $Z=0$,}
\end{equation}
\begin{equation}
A_{13}=\frac{3}{4}\omega_0B^2B^*_X-2i\alpha B_X-\frac{3\omega_0}{2}|B|^2B_X-\frac{\omega_0}{8k_0^4}B_{XXX}+\frac{i\omega_0}{g}B\bar{\phi}_{0T}+iB\bar{\phi}_{0X},\hspace*{1cm}\text{at $Z=0$.}
\end{equation}
\end{subequations}
Here $\Delta$ is the two-dimensional Laplacian operator, $\Delta F=F_{XX}+F_{ZZ}$.  
The first of the PDEs listed at each order can be solved as an advection equation with a complex velocity.  
The remaining PDEs are complex advection equations with nonhomogeneous terms 
that can be solved using variation of parameters.  
For brevity, the details have been omitted, 
but the solutions to these systems are included in the equations \eqref{surfvelpot}.

\section{Particle paths for generalizations of NLS}
We now present particle paths associated with three generalizations of the NLS equation.
We start with the higher-order non-viscous model, the Dysthe equation.
Then we continue with the dissipative nonlinear Schr\"{o}dinger equation (dNLS)
and finally with the viscous Dysthe equation.

\subsection{The Dysthe system}
\label{SectionMNLS}

\begin{figure}
\begin{center}
\includegraphics[width=12cm]{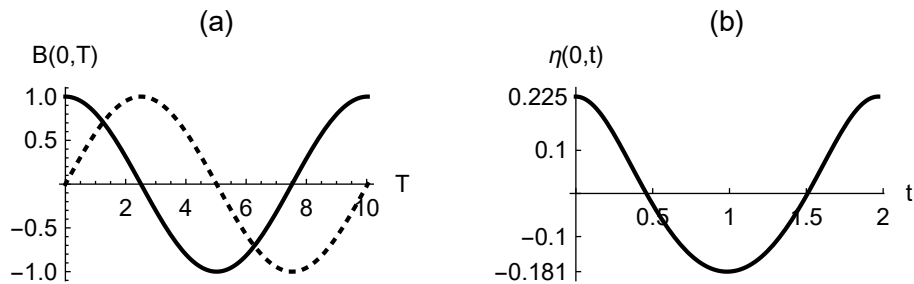}
\caption{\small Time series plots of (a) the real (--) and imaginary (- -) parts of the plane-wave solution to the Dysthe system, $B(0,T)$, and (b) the corresponding surface displacement  $\eta(0,t)$ for one period of the carrier wave for
$\epsilon=0.1$, $k_0=1$, $B_0=1$, and $k=0$.} 
\label{DysthePlots}
\end{center}
\end{figure}

\begin{figure}
\begin{center}
\includegraphics[width=6cm]{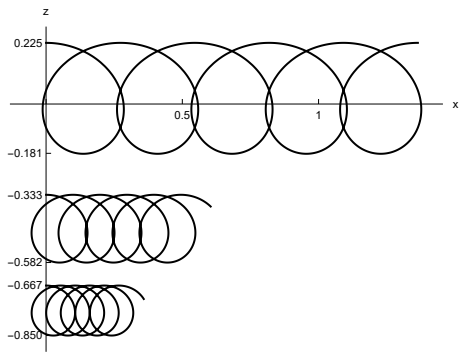}
\caption{\small The paths of three Dysthe plane-wave particles with $\epsilon=0.1$, $k_0=1$, $B_0=1$, and $k=0$ on $t\in[0,5T^*_{_{Dys}}]$.  The initial positions are $(0,0.2247)$, $(0,-0.5)$, and $(0,-1)$.} 
\label{NLS4LagPlots}
\end{center}
\end{figure}

\begin{figure}
\begin{center}
\includegraphics[width=12cm]{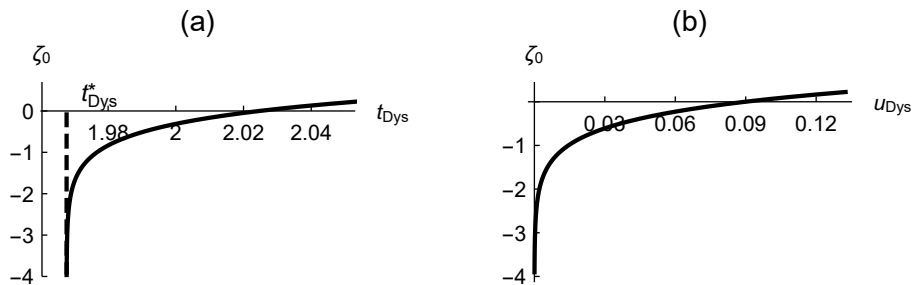}
\caption{\small Plots showing how (a) $t_{_{Dys}}$ and (b) $u_{_{Dys}}$ change for Dysthe plane-wave solutions as $\zeta_0$ changes.}
\label{NLS4LagData}
\end{center}
\end{figure}

The Dysthe system is obtained from Equation \eqref{vDysthe} by setting $\alpha=0$.  
This system has been shown to provide accurate predictions for the evolution of modulated wave trains 
for a wider range of $\epsilon$ values than NLS (see, for example, Lo \& Mei~\cite{LoMei}).  
Its plane-wave solutions are given by
\begin{equation}
B(X,T)=B_0\mbox{e}^{ikX-i\lambda_{_{Dys}}T},
\label{MNLSPWSoln}
\end{equation}
where
\begin{equation}
\lambda_{_{Dys}}=\omega_0\Big{(}\frac{k}{2k_0}+\frac{\epsilon
  k^2}{8k_0^2}-2\epsilon B_0^2k_0^2+\frac{\epsilon^2k^3}{16k_0^3}+5\epsilon^2B_0^2kk_0\Big{)},
\end{equation}
and $B_0$, $k$ are real constants.  The formulas for the corresponding surface displacement and velocity potential are included in Appendix \ref{etasMNLS}.  For demonstrative purposes, we select $\epsilon=0.1$, $B_0=1$, $k_0=1$ and $k=0$.  These parameters lead to a surface displacement with crest height of $0.2247$, trough height of $-0.1813$, and a $t$-period of 
\begin{equation}
t_{_{Dys}}^*=\frac{2\pi}{\omega_0-\epsilon\lambda_{Dys}}=\frac{100\pi\sqrt{5}}{357}\dot{=}1.968.
\end{equation}
This particular period is the same as in the NLS example examined above because we selected $k=0$ in both cases.  Figure \ref{DysthePlots} contains time-series plots of one period of the real and imaginary parts of the Dysthe plane-wave solution and one period of the corresponding surface displacement.  Figure \ref{NLS4LagPlots} shows the paths followed by three Dysthe plane-wave particles.  One starts on the surface and two start inside the fluid.  Figure \ref{NLS4LagData} contains plots demonstrating how the period and mean horizontal velocity depend on $\zeta_0$ for this Dysthe plane-wave solution.

For a given surface amplitude and $\zeta_0$, the periods, horizontal drifts, 
and average horizontal velocities of the Dysthe particles 
are all greater than the corresponding NLS quantities.  
The differences between the NLS and Dysthe results are small due to the orders of the equations,
and these differences go to zero as $z\rightarrow\infty$.  
Finally, for these Dysthe plane-wave solutions, the difference between the particle's vertical position 
and the fluid surface, as defined in Equation \eqref{error}, decreases like $\mathcal{O}(\epsilon^5)$.

\subsection{The dNLS equation}
\label{SectiondNLS}
\begin{figure}
\begin{center}
\includegraphics[width=6cm]{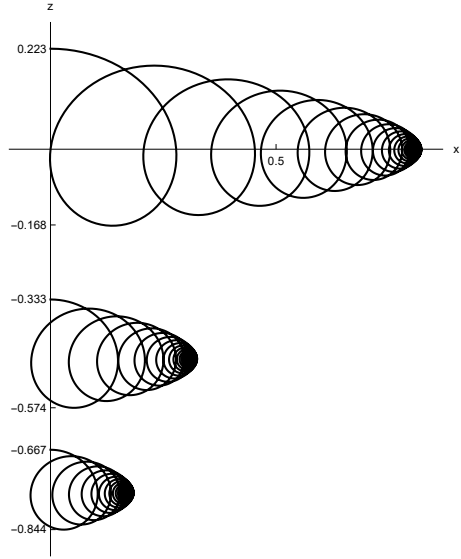}
\caption{\small The paths of three dNLS plane-wave particles with
  $\epsilon=0.10$, $B_0=1$, $k_0=1$, $k=0$, and $\alpha=4$.  The initial positions are $(0,0.223)$, $(0,-0.333)$, and
  $(0,-0.666)$.}
\label{dNLS3LagPlots}
\end{center}
\end{figure}

\begin{figure}
\begin{center}
\includegraphics[width=12cm]{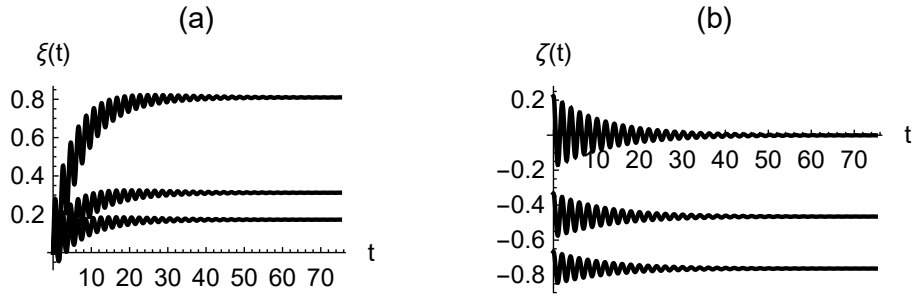}
\caption{\small Plots of (a) $\xi(t)$ and (b) $\zeta(t)$ for the three dNLS
  plane-wave particles in Figure \ref{dNLS3LagPlots}.}
\label{dNLS3xPlot}
\end{center}
\end{figure}

The dissipative nonlinear Schr\"odinger equation is obtained from \eqref{vDysthe} by setting $\epsilon^2=0$.  
The dNLS equation was first derived as a model of water waves by Dias {\emph{et al.}}~\cite{ddz08}.  
However, Lo \& Mei~\cite{LoMei} and Segur {\emph{et al.}}~\cite{bf05} had previously studied it 
and had shown that it accurately models the evolution of modulated wave trains.  
The plane-wave solutions of dNLS are given by
\begin{equation}
B(X,T)=B_0\mbox{exp}\Big{(}ikX+\omega_r(T)+i\omega_i(T)\Big{)},
\label{dNLSPWSoln}
\end{equation}
where
\begin{subequations}\label{dNLSPWs}
\begin{equation}
\omega_r(T)=-2\epsilon\alpha k_0^2T,
\end{equation}
\begin{equation}
\omega_i(T)=\omega_0\Big{(}-\frac{kT}{2k_0}-\frac{\epsilon
 k^2T}{8k_0^2}+\frac{B_0^2}{2\alpha}(1-\mbox{e}^{-4\epsilon\alpha k_0^2T})\Big{)},
\end{equation}
\end{subequations}
and $B_0$ and $k$ are real constants.  These solutions were chosen so that they limit to the plane-wave solutions of NLS as $\alpha\rightarrow0$.  Formulas for the corresponding surface displacement and velocity potential are included in Appendix \ref{etasMNLS}.  The plane-wave solutions of dNLS are not periodic in $T$, so the corresponding surface displacements are not periodic in $t$.  For demonstrative purposes, we selected $\epsilon=0.10$, $B_0=1$, $k_0=1$, $k=0$, and $\alpha=4$.  This value of $\alpha$ is roughly an order of magnitude larger than the experimentally determined parameters used by Segur {\emph{et al.}}~\cite{bf05} and Carter \& Govan~\cite{FD}.  We used a relatively large value for $\alpha$ so that viscous effects could be seen on relatively short time scales.  Figures \ref{dNLS3LagPlots} and \ref{dNLS3xPlot} show the paths of three dNLS plane-wave particles on the interval $t\in[0,75]$.  The top particle starts on the surface and the other two start and stay inside the fluid.  Although the motion of the particles is not periodic, a number of comments can still be made.  Due to dissipative effects, the motion, both horizontal and vertical, of all particles decreases as time increases.  Each particle eventually spirals in toward a fixed point.  The ``period'', the average horizontal Lagrangian velocity, and the horizontal drift all decrease as $t$ increases.  This is consistent with the NLS results where smaller-amplitude solutions lead to smaller periods, velocities, and drifts.

\begin{figure}
\begin{center}
\includegraphics[width=6cm]{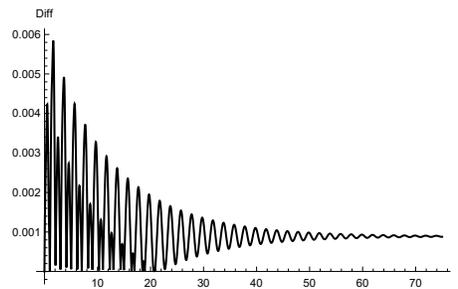}
\caption{\small A plot of $\mathcal{D}(t)$ versus $t$ corresponding to a dNLS
  plane-wave solution with $\epsilon=0.1$, $k_0=1$, $B_0=1$, $k=0$, and $(\xi_0,\zeta_0)=(0,0.223)$ on the interval
  $t\in[0,75]$.}
\label{dNLS3Diff}
\end{center}
\end{figure}

Figure \ref{dNLS3Diff} contains a plot of $\mathcal{D}(t)$ versus $t$ for this plane-wave solution of dNLS.  Note that $\mathcal{D}(t)$ does not limit to zero as $t\rightarrow\infty$.  This is because the particle that starts on the surface ends up inside the fluid.  This is due to two facts: (i) dNLS is an asymptotic model and (ii) the weakly viscous Euler equations are only an approximation to the true viscous system.  However, dNLS remains valid in the small viscosity, $\bar{\alpha}\rightarrow0$ limit.  Both $\mathcal{D}(t)$ and $\mathcal{E}(\epsilon)$ limit to zero as $\epsilon\rightarrow0$.

\subsection{The viscous Dysthe equation}
\label{SectionvDysthe}
\begin{figure}
\begin{center}
\includegraphics[width=6cm]{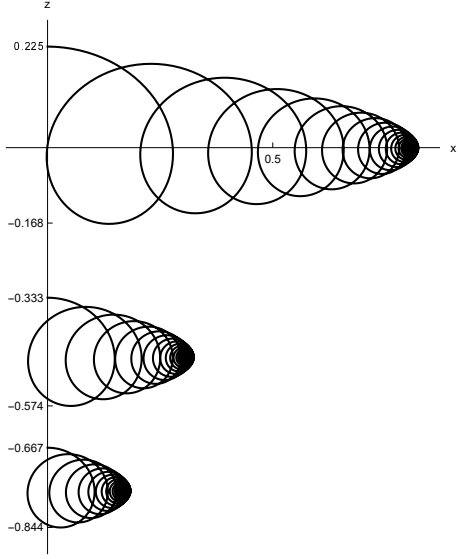}
\caption{\small The paths of three vDysthe plane-wave particles with $\epsilon=0.10$, $B_0=1$, $k_0=1$, $k=0$, and
  $\alpha=4$.  The initial positions are $(0,0.223)$, $(0,-0.333)$, and $(0,-0.666)$ on the interval $t\in[0,75]$.} 
\label{dNLS4LagPlots}
\end{center}
\end{figure}

\begin{figure}
\begin{center}
\includegraphics[width=12cm]{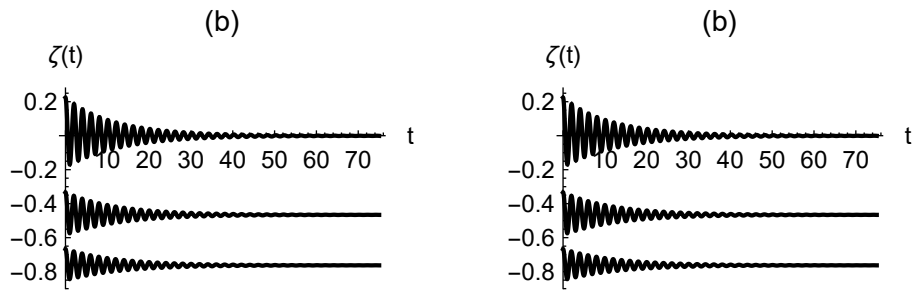}
\caption{\small Plots of (a) $\xi(t)$ and (b) $\zeta(t)$ for the three
  viscous Dysthe plane-wave particles in Figure \ref{dNLS4LagPlots}.}
\label{dNLS4xPlot}
\end{center}
\end{figure}

The viscous Dysthe system is given in Equation \eqref{vDysthe}.  
The plane-wave solutions of this system are given by
\begin{equation}
B(X,T)=B_0\mbox{exp}\Big{(}ikX+\omega_r(T)+i\omega_i(T)\Big{)},
\label{vDysthePWSoln}
\end{equation}
where
\begin{subequations}\label{vDysthePWs}
\begin{equation}
\omega_r(T)=-2\epsilon\alpha k_0(k0-2\epsilon k)T,
\end{equation}
\begin{equation}
\omega_i(T)=\omega_0\Big{(}\frac{kT}{2k_0}+\frac{\epsilon
 k^2T}{8k_0^2}+\frac{\epsilon^2k^3T}{16k_0^3}-\frac{B_0^2(2k_0-5\epsilon k)}{4\alpha(k_0-2\epsilon k)}\big{(}1-\mbox{e}^{2\omega_r(T)}\big{)}\Big{)},
\end{equation}
\end{subequations}
where $B_0$ and $k$ are real constants.  These solutions limit to the plane-wave solutions of the Dysthe system as $\alpha\rightarrow0$.  The plane-wave solutions of vDysthe are not periodic in $T$.  For demonstrative purposes, we selected $\epsilon=0.1$, $B_0=1$, $k_0=1$, $k=0$, and $\alpha=4$.  This value of $\alpha$ is roughly an order of magnitude larger than the experimentally determined parameters used by Segur {\emph{et al.}}~\cite{bf05} and Carter \& Govan~\cite{FD}.  We used a relatively large value for $\alpha$ so that viscous effects could be seen on relatively short time scales.  Figures \ref{dNLS4LagPlots} and \ref{dNLS4xPlot} show the paths of three vDysthe plane-wave particles on the interval $t\in[0,75]$.  There are many similarities with the dNLS case.  As $t$ increases, the ``period'', the average horizontal Lagrangian velocity, and the horizontal drift all decrease. The horizontal drift and average horizontal velocity limit to zero as $t\rightarrow\infty$.  The particles spiral in toward a fixed point.

\begin{figure}
\begin{center}
\includegraphics[width=6cm]{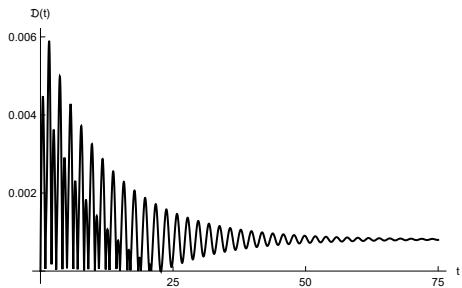}
\caption{\small A plot of $\mathcal{D}(t)$ versus $t$ corresponding to a vDysthe
  plane-wave solution with $\epsilon=0.1$, $k_0=1$,
  $B_0=1$, $k=0$, $\alpha=4$, and $(\xi_0,\zeta_0)=(0,0.223)$ over
  $t\in[0,75]$.}
\label{dNLS4Diff}
\end{center}
\end{figure}

Figure \ref{dNLS4Diff} shows that the error term $\mathcal{D}(t)$ 
is not smaller in the viscous Dysthe context than in the dNLS context.  
This observation further demonstrates the limitation that Equation \eqref{BC1} 
is only valid in the small $\alpha$ limit.  
Using a higher-order approximation to Equation \eqref{DDZ} does not provide a more accurate approximation 
to Equation \eqref{BC1} because Equation \eqref{BC1} 
with viscosity is only an approximation to the true viscous system.

\section{Conclusion}
In this paper, we used the classical derivation of the NLS equation to derive formulas for the velocity potential throughout the fluid.  Using these formulas, we numerically computed and examined the paths of fluid particles corresponding to plane-, cnoidal, and solitary-wave solutions to the NLS equation.  Following a similar procedure, we examined the paths of particles subject to the motion of plane-wave solutions to the Dysthe, dissipative NLS, and viscous Dysthe equations.  We showed that dissipative/viscous effects decrease particle speed and displacement. Finally, we showed that the boundary conditions of the full water wave problem are only asymptotically satisfied by the solutions to these equations.

\section*{Acknowledgments}
This research was supported by the Research Council of Norway under grant numbers 213474/F20 and 239300/F20, by the U.S.~National Science Foundation under grant number DMS-1716120, and by a Fulbright Core Scholar Award that allowed JDC to spend a semester visiting HK at the University of Bergen.  HK and JDC thank CIRM in Marseilles, France for hospitality during a two-week stay.  The authors also thank John Grue for helpful discussions.

\appendix

\section{Explicit formulas for surface displacement \& velocity potential}
\subsection{NLS plane-wave solutions}
\label{etasPWNLS}

The surface displacement and velocity potential for the NLS plane-wave solution 
given in Equation \eqref{NLSPWSoln} are
\begin{subequations}
\begin{equation}
\eta(x,t)=\epsilon B_0 \bar{E}+\epsilon^2 B_0^2k_0\bar{E}^2+\epsilon^3\Big{(}\frac{3}{2}B_0^3k_0^2\bar{E}^3-B_0^2k\bar{E}^2\Big{)}+c.c.,
\end{equation}
\begin{equation}
\phi(x,z,t)=\frac{i\epsilon B_0\omega_0}{k_0}\Big{(}1+\frac{\epsilon k}{2k_0}+\frac{3\epsilon^2k^2}{8k_0^2}-\frac{1}{2}\epsilon^2B_0^2k_0^2\Big{)}\tilde{E}+c.c.,
\end{equation}
\label{etaNLS}
\end{subequations}
where
\begin{subequations}
\begin{equation}
\bar{E}=\exp\Big{(}i\big{(}\omega_0-\epsilon\lambda_{_{NLS}}\big{)}t-i\big{(}k_0-\epsilon
k\big{)}x\Big{)},
\end{equation}
\begin{equation}
\tilde{E}=\bar{E}\exp\big{(}(k_0-\epsilon k)z\big{)}.
\end{equation}
\end{subequations}
Note that both the mean terms, $\bar{\eta}$ and $\bar{\phi}$, are zero for these solutions.

\subsection{NLS cnoidal-wave solutions}
\label{etasCNNLS}

The surface displacement and velocity potential for the NLS cnoidal-wave solution given 
in Equation \eqref{cnsoln} are
\begin{subequations}
\begin{equation}
\begin{split}
\eta(x,t)=&\epsilon B_0\mbox{cn}(r,\kappa)\bar{E}+\epsilon^2B_0^2k_0\mbox{cn}(r,\kappa)^2\bar{E}^2\\ &+\epsilon^3\Big{(}\frac{3}{2}B_0^3k_0^2\mbox{cn}(r,\kappa)^3\bar{E}^3-\frac{2\sqrt{2}i}{\kappa}B_0^3k_0^2\mbox{cn}(r,\kappa)\mbox{dn}(r,\kappa)\mbox{sn}(r,\kappa)\bar{E}^2\Big{)}\\ &+\epsilon^3\frac{\omega_0}{g}\big{(}\mathcal{H}(|B|^2)\big{)}_T+c.c.,
\end{split}
\end{equation}
\begin{equation}
\begin{split}
\phi(x,z,t)=&\frac{i}{k_0}\epsilon B_0\omega_0\mbox{cn}(s,\kappa)\tilde{E}+\epsilon^2\Big{(}-\frac{\sqrt{2}}{\kappa}B_0^2\omega_0\mbox{dn}(s,\kappa)\mbox{sn}(s,\kappa)\tilde{E}+\frac{\omega_0}{\pi}\int_{\mathbb{R}} i\mbox{sgn}(k)\mathcal{F}(\mbox{cn}^2(s,\kappa))\mbox{e}^{ikx+|k|z}dk\Big{)}\\ 
&+\epsilon^3\Big{(}\big{(}\frac{3i}{\kappa^2}B_0^3k_0\omega_0-6iB_0^3k_0\omega_0\big{)}\mbox{cn}(s,\kappa)\tilde{E}+\frac{11i}{2}B_0^3k_0\omega_0\mbox{cn}^3(s,\kappa)\tilde{E}\\ &\hspace*{1.3cm}+\frac{\sqrt{2}\omega_0B_0^3gk_0}{\kappa}\int_{\mathbb{R}}\mathcal{F}\big{(}\mbox{cn}(s,\kappa)\mbox{dn}(s,\kappa)\mbox{sn}(s,\kappa)\big{)}\mbox{e}^{ikx+|k|z} dk\Big{)}+c.c.,
\end{split}
\end{equation}
\end{subequations}
where
\begin{subequations}
\begin{equation}
r=\frac{\sqrt{2}}{\kappa}\epsilon B_0k_0\big{(}2k_0x-\omega_0t\big{)},
\end{equation}
\begin{equation}
s=r+\frac{2\sqrt{2}i}{\kappa}\epsilon B_0k_0^2z,
\end{equation}
\begin{equation}
\bar{E}=\exp\Bigg{(}i\omega_0\Big{(}1+2\epsilon^2 B_0^2k_0^2-\frac{\epsilon^2B_0^2k_0^2}{\kappa^2}\Big{)}t-ik_0x\Bigg{)},
\end{equation}
\begin{equation}
\tilde{E}=\bar{E}\exp(k_0z).
\end{equation}
\end{subequations}
Note that the mean terms are nonzero.  
\subsection{NLS solitary-wave solutions}
\label{etasSechNLS}

The surface displacement and velocity potential for the NLS solitary-wave solution given in 
Equation \eqref{sechsoln} are
\begin{subequations}
\begin{equation}
\begin{split}
\eta(x,t) = &  \epsilon B_0\sech(p)\bar{E}+\epsilon^2B_0^2k_0\sech^2(p)\bar{E}^2
                +\epsilon^3 \Big{(}\frac{3}{2}B_0^3k_0^2\sech^3(s)\bar{E}^3 + 
\\ &
 \qquad \qquad
2\sqrt{2}iB_0^3k_0^2\sech^2(p)\tanh(p)\bar{E}^2+\frac{\omega_0}{g}\big{(}\mathcal{H}(\sech^2(p)\big{)}_T\Big{)}+c.c.,
\end{split}
\end{equation}
\begin{equation}
\begin{split}
\phi(x,z,t) = & \frac{i}{k_0}\epsilon B_0\omega_0\sech(q)\tilde{E}
\\ &
 + \epsilon^2\Big{(}\sqrt{2}B_0^2\omega_0\sech(q)\tanh(q)\tilde{E}-\frac{\omega_0}{\pi}\int_{\mathbb{R}}i\mbox{sgn}(k)\mathcal{F}\big{(}\sech^2(q)\big{)}\mbox{e}^{-ikx+|k|z}dk\Big{)}
\\ &+\epsilon^3\Big{(}\frac{11i}{2}B_0^3k_0\omega_0\sech^3(q)\tilde{E}-3iB_0^3k_0\omega_0\sech(q)\tilde{E}
\\ &
\qquad \qquad
+ \sqrt{2}\omega_0B_0^3gk_0
  \int_{\mathbb{R}}\mathcal{F}\big{(}\mbox{sech}^2(q)\mbox{tanh}(q)\big{)}\mbox{e}^{ikx+|k|z} dk\Big{)}+c.c.,
\end{split}
\end{equation}
\end{subequations}
where
\begin{subequations}
\begin{equation}
p=\sqrt{2}\epsilon B_0k_0\big{(}2k_0x-\omega_0t\big{)},
\end{equation}
\begin{equation}
q=p-2\sqrt{2}i\epsilon B_0k_0^2z,
\end{equation}
\begin{equation}
\bar{E}=\exp\Big{(}i\omega_0\big{(}1+\epsilon^2B_0^2k_0^2\big{)}t-ik_0x\Big{)},
\end{equation}
\begin{equation}
\tilde{E}=\bar{E}\exp(k_0z).
\end{equation}
\end{subequations}
Note that the mean terms are nonzero.

\subsection{Dysthe plane-wave solutions}
\label{etasMNLS}

\begin{subequations}
\begin{equation}
\eta(x,t)=\epsilon B_0 \bar{E}+\epsilon^2B_0^2k_0\bar{E}^2+\epsilon^3\Big{({}\frac{3}{2}B_0^3k_0^2\bar{E}^3-B_0k\bar{E}^2\Big{)}+\epsilon^4\Big{(}\frac{17}{3}B_0^4k_0^3\bar{E}^2-3B_0^3kk_0\bar{E}^3+\frac{8}{3}B_0^4k_0^3\bar{E}^4}\Big{)}+c.c.,
\end{equation}
\begin{equation}
\phi(x,z,t)=\frac{i\epsilon B_0\omega_0}{k_0}\Big{(}1+\frac{\epsilon k}{2k_0}+\frac{3\epsilon^2k^2}{8k_0^2}-\frac{1}{2}\epsilon^2B_0^2k_0^2+\frac{3}{4}\epsilon^3B_0^2kk_0+\frac{5\epsilon^3k^3}{16k_0^3}\Big{)}\tilde{E}+4\epsilon^4iB_0^4\omega_0\tilde{E}^2k_0^2+c.c.,
\end{equation}
\label{etaMNLS}
\end{subequations}
where
\begin{subequations}
\begin{equation}
\bar{E}=\exp\Big{(}i\big{(}\omega_0-\lambda_{_{Dys}}\big{)}t-i\big{(}k_0-\epsilon
k\big{)}x\Big{)},
\end{equation}
\begin{equation}
\tilde{E}=\bar{E}\exp\Big{(}\big{(}k_0-\epsilon k\big{)}z\Big{)}.
\end{equation}
\end{subequations}
As expected, the Dysthe surface displacement and velocity potential up to $\mathcal{O}(\epsilon^3)$ are the same as those in NLS.

\subsection{dNLS plane-wave solutions}
\label{etasdNLS}

\begin{subequations}
\begin{equation}
\eta(x,t)=\epsilon B_0 \bar{E}+\epsilon^2 B_0^2k_0\bar{E}^2+\epsilon^3\Big{(}\frac{3}{2}B_0^3k_0^2\bar{E}^3-B_0^2k\bar{E}^2\Big{)}+c.c.,
\end{equation}
\begin{equation}
\phi(x,z,t)=\frac{i\epsilon B_0\omega_0}{k_0}\Big{(}1+\frac{\epsilon k}{2k_0}+\frac{3\epsilon^2k^2}{8k_0^2}\Big{)}\tilde{E}-\frac{1}{2}i\epsilon^3B_0^3k_0\omega_0\tilde{E}\mbox{e}^{-4\epsilon^2\alpha k_0^2t}+c.c.,
\end{equation}
\label{etadNLS}
\end{subequations}
where
\begin{subequations}
\begin{equation}
\bar{E}=\exp\Big{(}i\omega_0t+\omega_r(\epsilon t)+i\omega_i(\epsilon t)-i\big{(}k_0-\epsilon
k\big{)}x\Big{)}.
\end{equation}
\begin{equation}
\tilde{E}=\bar{E}\exp{\Big{(}\big{(}k_0-\epsilon k\big{)}x\Big{)}},
\end{equation}
\end{subequations}
where $\omega_r$ and $\omega_i$ are given in Equation \eqref{dNLSPWs}.
Note that the mean terms are zero.  Also note that the nonlinear term in dNLS causes some portions of the velocity potential to decay faster than others. 
\subsection{Viscous Dysthe plane-wave solutions}
\label{etasvDysthe}

\begin{subequations}
\begin{equation}
\begin{split}
\eta(x,t)=&\epsilon B_0 \bar{E}+\epsilon^2B_0^2k_0\bar{E}^2+\epsilon^3\Big{({}\frac{3}{2}B_0^3k_0^2\bar{E}^3-B_0k\bar{E}^2\Big{)}+\epsilon^4\Big{(}\frac{4i}{g}B_0^2\alpha k_0^2\omega_0\bar{E}^2-3B_0^3kk_0\bar{E}^3}\Big{)}\\ &+\epsilon^{4}\Big{(}\frac{17}{3}B_0^4k_0^3\bar{E}^2\mbox{e}^{-4\epsilon^2k_0\alpha(k_0-2\epsilon k)t}+\frac{8}{3}B_0^4k_0^3\bar{E}^4\mbox{e}^{-8\epsilon^2k_0\alpha(k_0-2\epsilon k)t}+\Big{)}+c.c.,
\end{split}
\end{equation}
\begin{equation}
\begin{split}
\phi(x,z,t)=&\frac{i\epsilon B_0\omega_0}{k_0}\Big{(}1+\frac{\epsilon k}{2k_0}+\frac{3\epsilon^2k^2}{8k_0^2}+\frac{5\epsilon^3k^3}{16k_0^3}\Big{)}\tilde{E}+\epsilon^4\Big{(}-\frac{1}{2}i\epsilon^3B_0^3k_0\omega_0+\frac{3}{4}i\epsilon^4B_0^3k\omega_0\Big{)}\tilde{E}\mbox{e}^{-4\epsilon^2k_0(k_0-2\epsilon k)t}\\ & +4\epsilon^4iB_0^4k_0^2\omega_0\tilde{E}^2\mbox{e}^{-8\epsilon^2k_0(k_0-2\epsilon k)t}-2\epsilon^4B_0^2\alpha k_0^2\tilde{E}^2+c.c.,
\end{split}
\end{equation}
\label{etavDysthe}
\end{subequations}
where
\begin{subequations}
\begin{equation}
\bar{E}=\exp\Big{(}i\omega_0t+\omega_r(\epsilon t)+i\omega_i(\epsilon t)-i\big{(}k_0-\epsilon
k\big{)}x\Big{)}.
\end{equation}
\begin{equation}
\tilde{E}=\bar{E}\exp{\Big{(}\big{(}k_0-\epsilon k\big{)}x\Big{)}},
\end{equation}
\end{subequations}
where $\omega_r$ and $\omega_i$ are given in Equation \eqref{vDysthePWs}.  
Just as with all the other plane-wave solutions, the mean terms here are zero.

\bibliographystyle{plain}

\end{document}